\documentclass[fleqn,usenatbib]{mnras}

\usepackage{newtxtext,newtxmath}
\usepackage{anyfontsize}

\usepackage[T1]{fontenc}

\DeclareRobustCommand{\VAN}[3]{#2}
\let\VANthebibliography\thebibliography
\def\thebibliography{\DeclareRobustCommand{\VAN}[3]{##3}\VANthebibliography}

\usepackage{graphicx}	
\usepackage{amsmath}	
\usepackage{xcolor}
\usepackage{subcaption}
\usepackage{siunitx}
\usepackage[normalem]{ulem}

\title[Optical variations in Swift~J1858.6--0814]{Long term optical variations in Swift~J1858.6--0814: evidence for ablation and comparisons to radio properties}

\author[Rhodes et al. 2024]{L. Rhodes$^{1}$\thanks{E-mail: lauren.rhodes@physics.ox.ac.uk},
D. M. Russell$^{2}$,
P. Saikia$^{2}$,
K. Alabarta$^{2}$,
J. van den Eijnden$^{3}$,
A. H. Knight$^{4}$,
M. C. Baglio$^{5}$,\newauthor
F. Lewis$^{6,7}$
\\
$^{1}$Astrophysics, Department of Physics, University of Oxford, Denys Wilkinson Building, Keble Road, Oxford OX1 3RH, UK\\
$^{2}$Center for Astrophysics and Space Science (CASS), New York University Abu Dhabi, PO Box 129188, Abu Dhabi, UAE\\
$^{3}$Department of Physics, University of Warwick, Coventry CV4 7AL, UK\\
$^{4}$Centre for Extragalactic Astronomy, Department of Physics, Durham University, South Road, Durham DH1 3LE, UK\\
$^{5}$INAF–Osservatorio Astronomico di Brera, Via Bianchi 46, I-23807 Merate (LC), Italy\\
$^{6}$Faulkes Telescope Project, School of Physics and Astronomy, Cardiff University, The Parade, Cardiff, CF24 3AA, Wales, UK\\
$^{7}$The Schools' Observatory, Astrophysics Research Institute, Liverpool John Moores University, 146 Brownlow Hill, Liverpool L3 5RF, UK
}

\date{Accepted XXX. Received YYY; in original form ZZZ}

\pubyear{2023}

\begin{document}
\label{firstpage}
\pagerange{\pageref{firstpage}--\pageref{lastpage}}
\maketitle

\begin{abstract}
We present optical monitoring of the neutron star low-mass X-ray binary Swift J1858.6–0814 during its 2018–2020 outburst and subsequent quiescence. We find that there was strong optical variability present throughout the entire outburst period covered by our monitoring, while the average flux remained steady. 
The optical spectral energy distribution is blue on most dates, consistent with emission from an accretion disc, interspersed by occasional red flares, likely due to optically thin synchrotron emission. We find that the fractional rms variability has comparable amplitudes in the radio and optical bands. This implies that the long-term variability is likely to be due to accretion changes, seen at optical wavelengths, that propagate into the jet, seen at radio frequencies. We find that the optical flux varies asymmetrically about the orbital period peaking at phase $\sim$0.7, with a modulation amplitude that is the same across all optical wavebands suggesting that reprocessing off of the disc, companion star and ablated material is driving the phase dependence. The evidence of ablation found in X-ray binaries is vital in understanding the long term evolution of neutron star X-ray binaries and how they evolve into (potentially isolated) millisecond pulsars.
\end{abstract}




\section{Introduction} 

Low-mass X-ray binaries (LMXBs) are binary systems made up of a compact object, either a black hole (BH) or neutron star (NS), and a low-mass stellar companion. Mass is transferred from the star to the compact object via Roche Lobe overflow through the inner Lagrange point. A given LMXB can go through a period of increased mass accretion rate called an `outburst', producing luminous, variable emission from radio to X-ray energies, which can last between weeks and months \citep[see e.g. ][for a review]{2023hxga.book..120B}. The earliest sign of an LMXB going into outburst is thought to be a sharp increase in optical flux \citep[see][and references therein]{2019AN....340..278R}. For BH LMXBs, near the beginning of the outburst, there is also a radio brightening accompanying an X-ray rise. During the hard state, named after the hard, power-law X-ray spectrum, any radio emission is thought to originate from a compact jet \citep{2001MNRAS.322...31F}. As the outburst evolves and the system transitions to a more X-ray-luminous, disc-dominated `soft state', the radio emission is quenched \citep*{2004MNRAS.355.1105F}. However, for XRBs hosting NSs, this process is more complex. For example, the radio emission is not fully quenched when NS XRBs transition into soft state \citep[see ][for examples of radio detections and deep limits in the soft state, respectively]{2004MNRAS.351..186M, 2017MNRAS.470.1871G}. The origin of the optical emission in LMXBs is more complex because the total flux may be a combination of multiple components. During quiescence, the optical counterpart can be dominated by the companion star in the system, or by low level accretion activity \citep*[e.g.][]{Greene2001,Zurita2003,Cantrell2010,Baglio2017,2022ApJ...930...20B}. An associated jet contribution is seen rarely. \citep{Baglio2013,Plotkin2016,RussellAlQasim2018}. During outbursts, the observed emission is thought to be mostly a superposition of optical emission from the disc, reprocessed X-ray emission on the disc surface, and a jet component. Multi-frequency optical data is required to interpret whether the emission is from the disc or jet as the two components are blue and red in colour, respectively \citep*[e.g.][]{Corbel02,Greenhill2006,Russell2007,Rahoui2012}. In addition to the long-term flux changes, many LMXBs also show shorter-term (sub-seconds to days) variability, corresponding to rapid changes in accretion rate, quasi-periodic oscillations from the inner accretion flow, flickering from the jet or bursting from the neutron star surface and inner accretion flow \citep[e.g.][]{Casella2010,veledina2011,2015MNRAS.449..268B,Gandhi2017,Malzac2018,IngramMotta2019,Tetarenko2019}.

NS LMXBs are thought to be progenitors to millisecond pulsars (MSPs), as confirmed by the discovery of the transitional MSP PSR J1023+0038 \citep{1982Natur.300..728A, 1998Natur.394..344W,Archibald2009,2013Natur.501..517P}. Some MSPs in binary systems are known as `spider pulsars' thanks to evidence of ablation of the stellar companion by the pulsar wind. It may be possible for complete ablation of the stellar companion to occur resulting in an isolated MSP \citep{Roberts2013}. More recently, evidence of ablation has been found in NS LMXBs whilst in outburst where ablation is thought to be driven by X-ray irradiation rather than the pulsar wind, thus suggesting that ablation is a more common process in NS binary systems than was originally thought, including a recently discovered system \textit{Swift} J1858.6--0814 \citep{2023MNRAS.520.3416K}.

\subsection{\textit{Swift} J1858.6--0814}

\textit{Swift} J1858.6--0814 (hereafter J1858) was first reported by the Neil Gehrels \textit{Swift} Observatory - Burst Alert Telescope (BAT) as a Galactic X-ray transient on 2018 October 25 \citep[MJD 58416, ][ the dotted vertical line in Figure \ref{fig:optical_radio_lightcurve}]{Krimm2018}. The UV, optical and radio counterparts were soon identified \citep{Kennea2018, Vasilopoulos2018, 2018ATel12184....1B}. The source remained in a highly variable, flaring state until February 2020, the system underwent a state transition \citep[the vertical dashed line in Figure \ref{fig:optical_radio_lightcurve} ][]{2020ATel13536....1B} finally by May 2020 (MJD 58970) J1858 was in a quiescent state \citep{2020ATel13719....1S, 2020ATel13725....1P}.

Given its highly variable nature, J1858 has been the subject of many in-depth campaigns at X-ray energies. We highlight two findings of interest for their relevance for the work presented here: (1) type-I X-ray bursts and (2) X-ray eclipses. The former confirmed the nature of the compact object as a neutron star and the latter provided strong constraints on the system inclination of $\sim81^{\circ}$ \citep{2020MNRAS.499..793B,2021MNRAS.503.5600B, 2022MNRAS.514.1908K,2023MNRAS.520.3416K, Vincentelli2023}. \citet{2022MNRAS.514.1908K} also found evidence of ablation of the companion star, i.e. the removal of material from the surface of the companion star mostly likely by high energy radiation from the inner accretion flow, analogous to the pulsar wind-driven ablation found in spider pulsars \citep{2001MNRAS.321..576S}. 

There have been a number of targeted short-timescale variability studies of J1858 across all wavelength bands \citep[e.g.][]{CastroSegura2022, shabaz2023, vandenEijnden2020}. The long-term radio behaviour of J1858 is best described by an initial flare, followed by a variable, persistent source that remained until the system underwent a state transition. The radio source was attributed to a compact jet, determined by a flat/positive ($\alpha > 0$, Flux density $\propto \nu^{\alpha}$) spectral index \citep{Rhodes2022, vandenEijnden2020}. On shorter timescales of $\sim$ minutes, the radio emission was highly variable with root mean square (rms) variability measurements ranging between 15 and 60 per cent across the outburst. The rapid variability was accompanied by a broader range of spectral index measurements swapping between optically thick and optically thin synchrotron spectra \citep{Rhodes2022, Vincentelli2023}. The optically thick-thin transitions were interpreted as successive jet ejections as opposed to the expected canonical hard state jet.

At optical wavelengths, the emission is dominated by the accretion disc. Perhaps the most interesting characteristic of the optical data published thus far is the rapid variability as reported by \citet{shabaz2023}. Similar behaviour was also reported by \citet{2018ATel12180....1B,2018ATel12186....1R,2018ATel12197....1P} and \citet{Vincentelli2023}. The flaring behaviour is split into `blue' and `red' flares. The red flares are interpreted as originating from the jet, they are shorter in duration (10s of seconds) and smaller in amplitude. However, as of yet no direct comparisons have been made with the radio properties of the system. The brighter, longer blue flares (100-400\,seconds) are consistent with an accretion disc spectrum \citep{MunozDarias2020,CastroSegura2022}.

In this paper, we present new long-term optical monitoring of J1858 from the Las Cumbres Observatory and discuss the variability properties of the source compared to what has been observed at radio frequencies. Section \ref{sec:obs} lays out the optical observing campaign and recaps the radio monitoring programs for J1858. In Section \ref{sec:results}, we present the results of our analysis of the optical data in both time and frequency space, comparing the optical and radio variability and spectral properties, before interpreting them within the picture of NS LMXBs and spider pulsars. Finally, in Section \ref{sec:conclusions} we lay out our conclusions.

\section{Observations} \label{sec:obs}

\subsection{Faulkes Telescope / LCO monitoring} \label{sec:obsLCO}

We began monitoring \textit{Swift} J1858.6--0814 with the 1-m and 2-m Las Cumbres Observatory \citep[LCO;][]{Brown2013} network of optical telescopes in November 2018, starting with some fast-timing observations with the 2-m Faulkes Telescope North (at Haleakala Observatory, Maui, Hawai`i, USA) on 2018 November 6 \citep[initial results were presented in][]{2018ATel12180....1B}. Monitoring continued throughout the outburst using both 2-m Faulkes Telescopes (Faulkes Telescope South is located at Siding Spring Observatory, Australia) and the 1-m network, which includes nodes at Siding Spring Observatory (Australia), Cerro Tololo Inter-American Observatory (Chile), McDonald Observatory (Texas, USA), Teide Observatory (Tenerife, Canary Islands, Spain) and the South African Astronomical Observatory (SAAO, South Africa). The LCO network comprise of robotic telescopes optimized for research and education \citep[e.g.][]{Lewis2018}.

Imaging was carried out roughly weekly in the SDSS $g^{\prime}$, $r^{\prime}$, $i^{\prime}$ and PanSTARRS $Y$-band filters during the outburst, and continued in the $i^{\prime}$-band in quiescence after the outburst, as part of an on-going monitoring campaign of $\sim$50 LMXBs \citep{Lewis2008} coordinated by the Faulkes Telescope Project. Exposure times were between 60 s and 200 s. In addition, a sequence of 50 $r^{\prime}$-band exposures were taken on 2018-11-06, with exposure times of 30 s \citep[the time resolution was 46 s;][]{2018ATel12180....1B}.
Data reduction and aperture photometry were carried out by the ``X-ray Binary New Early Warning System (XB-NEWS)'' data analysis pipeline \citep{2019AN....340..278R,Goodwin2020}. The pipeline downloads images and calibration data from the LCO archive, performs several quality control steps to reject any bad quality images, and computes an astrometric solution on each image using Gaia DR2\footnote{\url{https://www.cosmos.esa.int/web/gaia/dr2}} positions, and then performs aperture photometry of all the stars. The method described in \citet{Bramich2012} is used to solve for zero-point magnitude offsets between epochs, and multi-aperture photometry is used on the target. Flux calibration of all stars is achieved using the ATLAS All-Sky Stellar Reference Catalog \citep[ATLAS-REFCAT2,][]{Tonry2018}\footnote{\url{https://archive.stsci.edu/prepds/atlas-refcat2/}}, which includes APASS, PanSTARRS DR1, and other catalogues to extract the magnitudes of the source. When the source is not detected above the detection threshold by the pipeline, XB-NEWS performs forced photometry at the known location of the source. Magnitudes with errors $> 0.25$ mag are excluded as these are either very marginal detections or non-detections. Since XB-NEWS uses multi-epoch detections to derive a best-fit position for the source, we report here the coordinates derived of \textit{Swift} J1858.6--0814 by XB-NEWS, of RA = 18\:58\:34.905, Dec = -08\:14\:14.94 (J2000; with an error of less than $\sim 0.2$\,arcsecond from catalogue cross-calibration uncertainties). This is consistent (within 1\,arcsecond) with the \textit{Swift} UVOT position \citep{Kennea2018} and the PanSTARRS position (within 0.2\,arcsecond; object ID 98112846453925483; observed during quiescence).

\subsection{Radio data} 

We use radio observations presented in \citet{vandenEijnden2020} and \citet{Rhodes2022}, to contextualise the optical data described in Section \ref{sec:obsLCO}. The data from these two publications cover both long-term monitoring campaigns and short-term variability studies. The data we use from these studies covers three frequencies: 1.28\,GHz (MeerKAT), 4.5\,GHz (the VLA) and 15.5\,GHz (AMI-LA). Variability was observed on both long (week-month) and short (minute) timescales. Across the observing campaigns, the radio spectrum was consistent with optically thick emission i.e. a steep spectrum ($\alpha >0$).

\section{Results} \label{sec:results}

\begin{figure*}
    \centering
    \includegraphics[width = \textwidth]{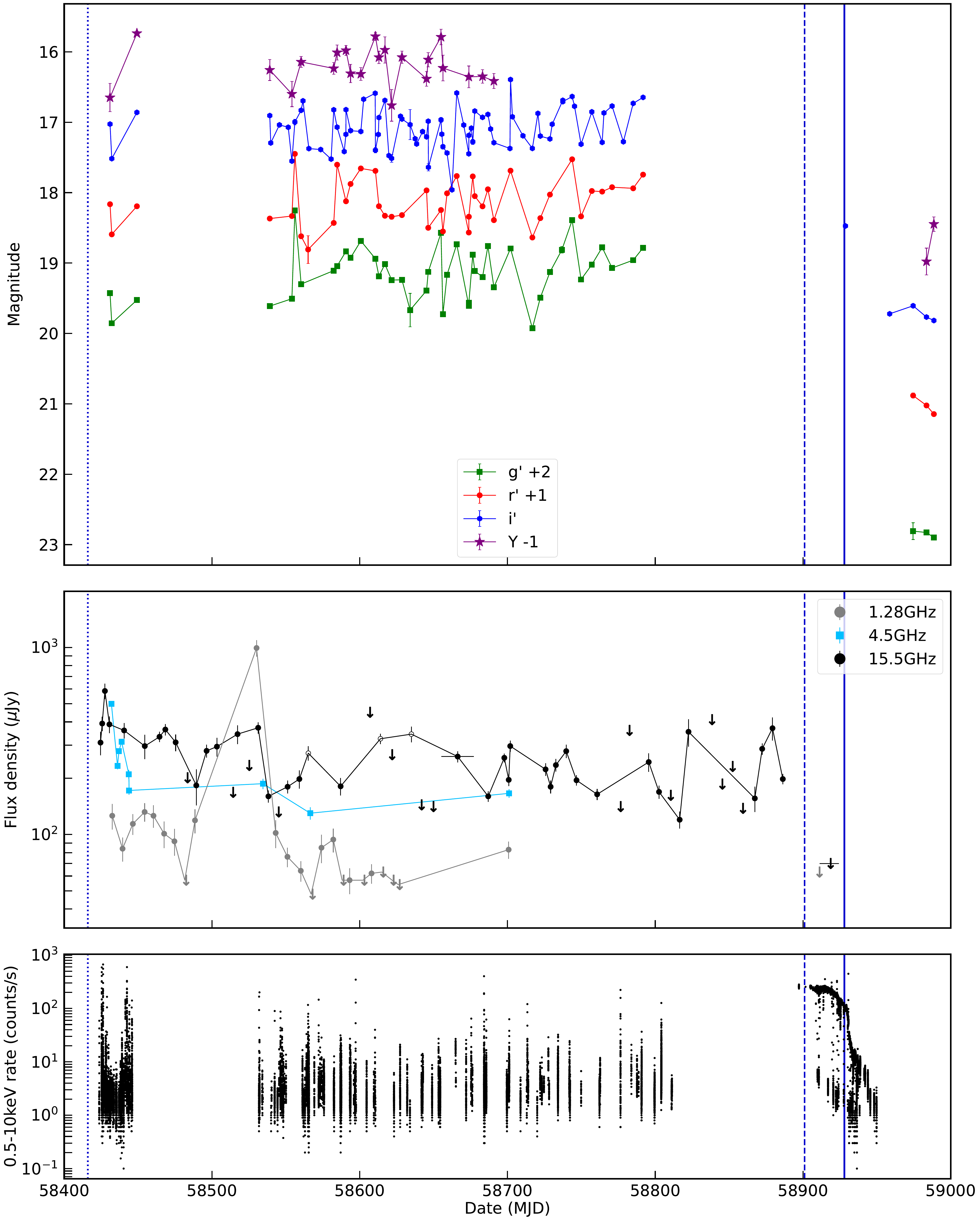}
    \caption{\textit{Upper panel:} Optical light curves of the J1858 system during outburst and quiescence. Artificial offsets have been placed to make the data easier to visualise. The vertical dotted, dashed and solid blue lines indicate the outburst's beginning, state transition and return to quiescence, respectively. \textit{Middle panel:} The radio data from \citet{vandenEijnden2020, Rhodes2022}. \textit{Lower panel:} The 0.5-10\,keV NICER light curve from \citet{CastroSegura2022}.}
    \label{fig:optical_radio_lightcurve}
\end{figure*}

\subsection{Light curves}
\label{subsec:long_term_optical}

Figure \ref{fig:optical_radio_lightcurve} shows a summary of J1858's outburst through its multi-wavelength light curves. Across all wavebands there is significant and rapid variability. The upper panel of Figure \ref{fig:optical_radio_lightcurve} shows the optical data of J1858 from our observing campaign starting from MJD 58431 (November 9\textsuperscript{th} 2018) and continuing until MJD 60126 (May 19\textsuperscript{th} 2020). The central panel in Figure \ref{fig:optical_radio_lightcurve} shows the long-term radio monitoring from \citet{vandenEijnden2020, Rhodes2022}. The lower panel shows the 0.5-10\,keV NICER X-ray light curve from \citet{CastroSegura2022}. In the X-ray light curve the transition from the flaring to steady state (denoted by the vertical dashed line) and the fade into quiescence (the solid vertical line) are both very clear.

X-ray binary optical light curves typically follow a the typical fast rise, exponential decay trend \citep[e.g.][]{2013MNRAS.432.1133M, 2020MNRAS.498.3429G, 2021MNRAS.504..444C,2023MNRAS.524.4543S}. In the case of J1858, there are no clear long-term trends, the optical counterpart in the upper panel of Figure \ref{fig:optical_radio_lightcurve}, has a mean magnitude in $g^{\prime}$, $r^{\prime}$, $i^{\prime}$ and $y$-bands of 17.1$\pm$0.4, 17.1$\pm$0.3, 17.1$\pm$0.3 and 17.2$\pm$0.3, respectively, (not corrected for interstellar extinction); the uncertainties are 1$\sigma$ and for comparison, the average uncertainty on a single data point is 0.04\,mag. The uncertainty on the average magnitude reflects the significant variability in each band. We find that the amplitude of variability is strongly correlated with frequency, such that the highest frequency has the highest amplitude and the lowest frequency has the lowest amplitude. The fractional rms (F\textsubscript{rms}) variability \citep{vaughan} across our 400\,day-long observing campaign of $\sim$weekly observations is $19\pm3$\% in $y$-band and $35.5\pm0.6$\% in $g^{\prime}$-band.

\begin{figure}
    \centering
    \includegraphics[width = \columnwidth]{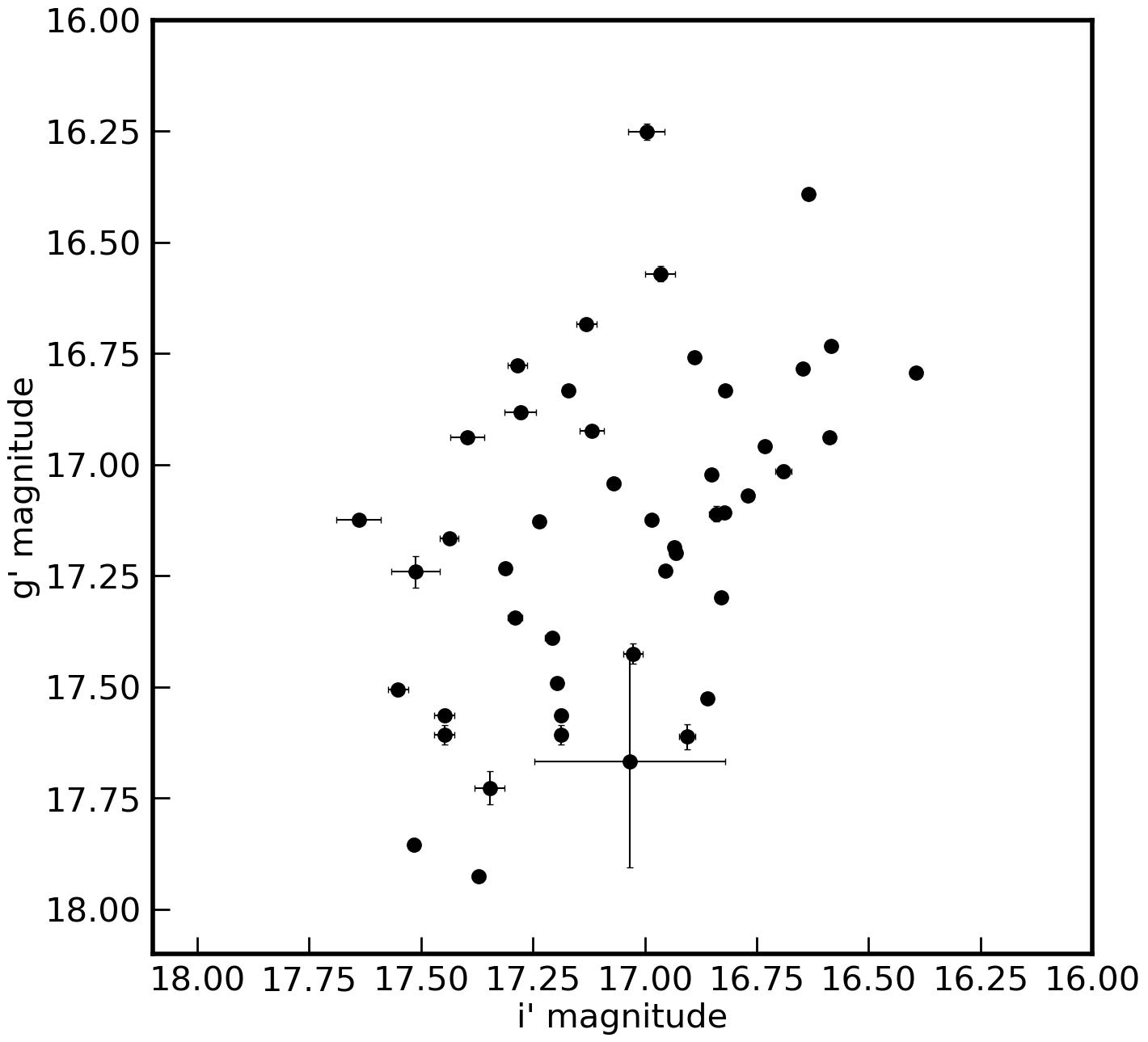}
    \caption{$g^{\prime}$ vs $i^{\prime}$ magnitude for observations made within the same night (most are taken within a few minutes during a filter sequence). There is evidence of a correlation but with considerable scatter, indicating that the variability timescale is shorter than the time between observations in different filters.}
    \label{fig:g_vs_i}
\end{figure}

\begin{figure}
\centering
\includegraphics[width = \columnwidth]{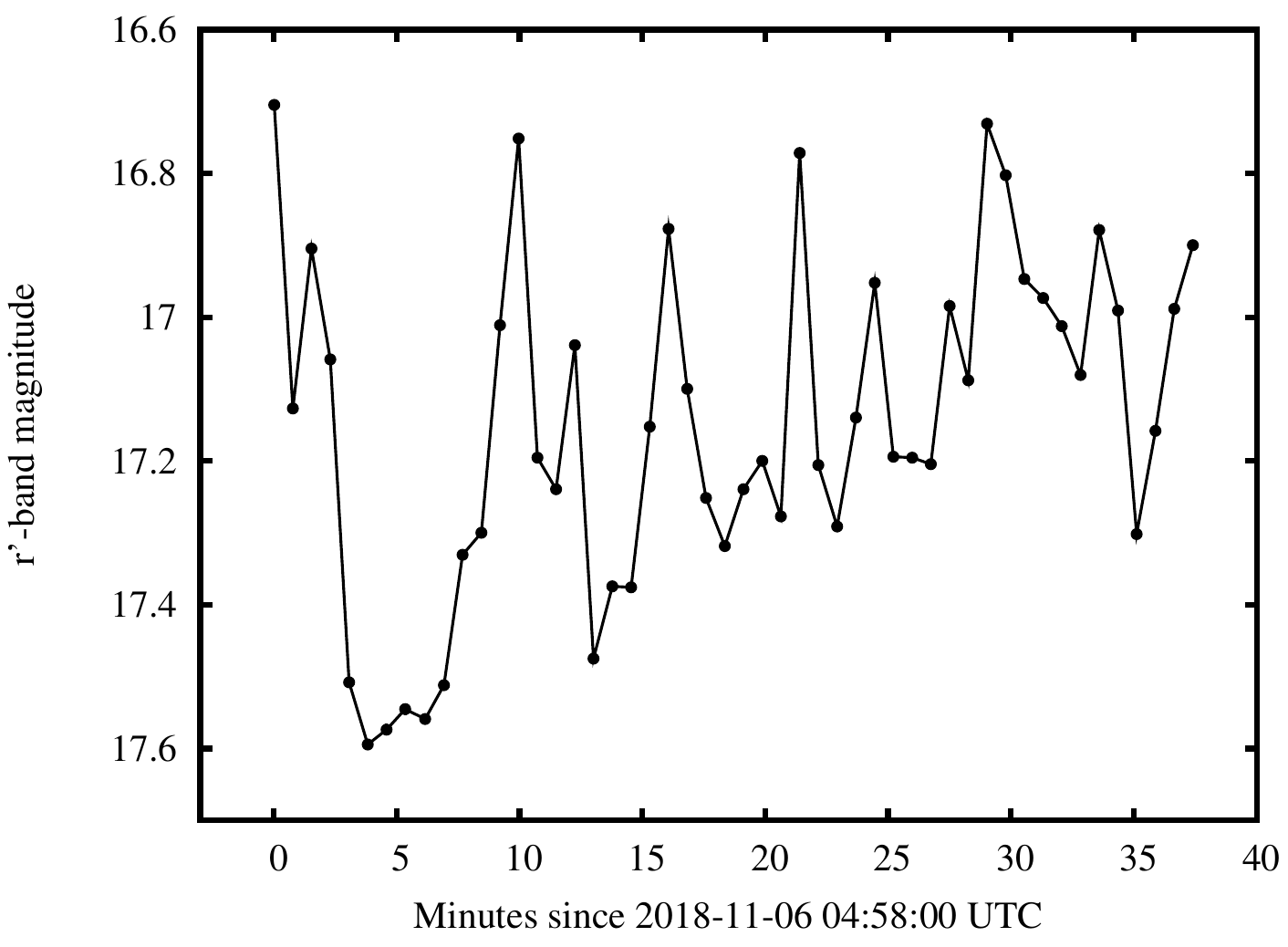}
\caption{Short-term optical variability observed near the start of the outburst, demonstrating that the source can vary as much as 0.9 magnitudes on timescales of a few minutes. Magnitude errors are plotted, but are generally smaller than the symbols. Flares and dips on these timescales are similar to those reported in \citet{MunozDarias2020}, \citet{Vincentelli2023} and \citet{shabaz2023}.} 
\label{fig:r-band-timing}
\end{figure}

Between MJD 58700 and 58750, there are potential signs of correlated behaviour across the $g^{\prime}$, $r^{\prime}$ and $i^{\prime}$-bands. However, across the outburst, we find only weakly correlated behaviour between all or any two given bands (as demonstrated by the large scatter in Figure \ref{fig:g_vs_i}). This suggests that the intrinsic variability timescale is shorter than the time taken to observe in two successive bands. This is further demonstrated in Figures  \ref{fig:g_vs_i} and \ref{fig:r-band-timing}. In Figure \ref{fig:g_vs_i}, for a given $g^{\prime}$ or $i^{\prime}$-band measurement, there is about one magnitude of scatter in the other band. 
Figure \ref{fig:r-band-timing} shows the results of a fast photometry observation (36-s time resolution; Table 1) with the 2-m Faulkes Telescope North \citep[see also][]{2018ATel12180....1B}. We observe flaring behaviour on minute timescales where the observed flux varied by as much as 0.9\,magnitudes which is consistent with the findings in Figure \ref{fig:g_vs_i}.

Around MJD 58928, J1858 fades towards quiescence \citep{2020ATel13719....1S}, during which time, we only have observations in $r^{\prime}$, $i^{\prime}$ and $y$-bands. In quiescence, the source is between 2.5 and 3 magnitudes fainter than during outburst, considering only the epochs with the best seeing (i.e. no contamination from the contaminating star) we measure the average magnitudes of 20.85$\pm$0.04, 20.02$\pm$0.04, 19.72$\pm$0.08 and 19.7$\pm$0.3 in the $g^{\prime}$, $r^{\prime}$, $i^{\prime}$ and $y$ bands, respectively.

The middle panel in Figure \ref{fig:optical_radio_lightcurve} shows the long-term radio monitoring from \citet{vandenEijnden2020, Rhodes2022}. Similarly to the optical data, we observe significant variability. We measure rms variability values of 20$\pm$4\%, 46$\pm$10\% and 33$\pm$2\% at 1.28, 4.5 and 15.5\,GHz, respectively, on week long timescales. The variability observed at 4.5\,GHz appears to be mildly higher, however, we note there are also larger uncertainties; both of these may be the result of a concentrated number of data points at the beginning of the outburst but a low number in total. Like the optical data, we find that the lowest radio frequencies have the smallest F\textsubscript{rms} values. The levels of variability in the radio band are consistent with that at optical wavelengths, indicating a possible common origin, or related components. After the state transition, denoted by the vertical blue dashed line, we obtain deep radio limits and only a single optical observation was made during the decay into quiescence.

\begin{figure}
    \centering
    \includegraphics[width = \columnwidth]{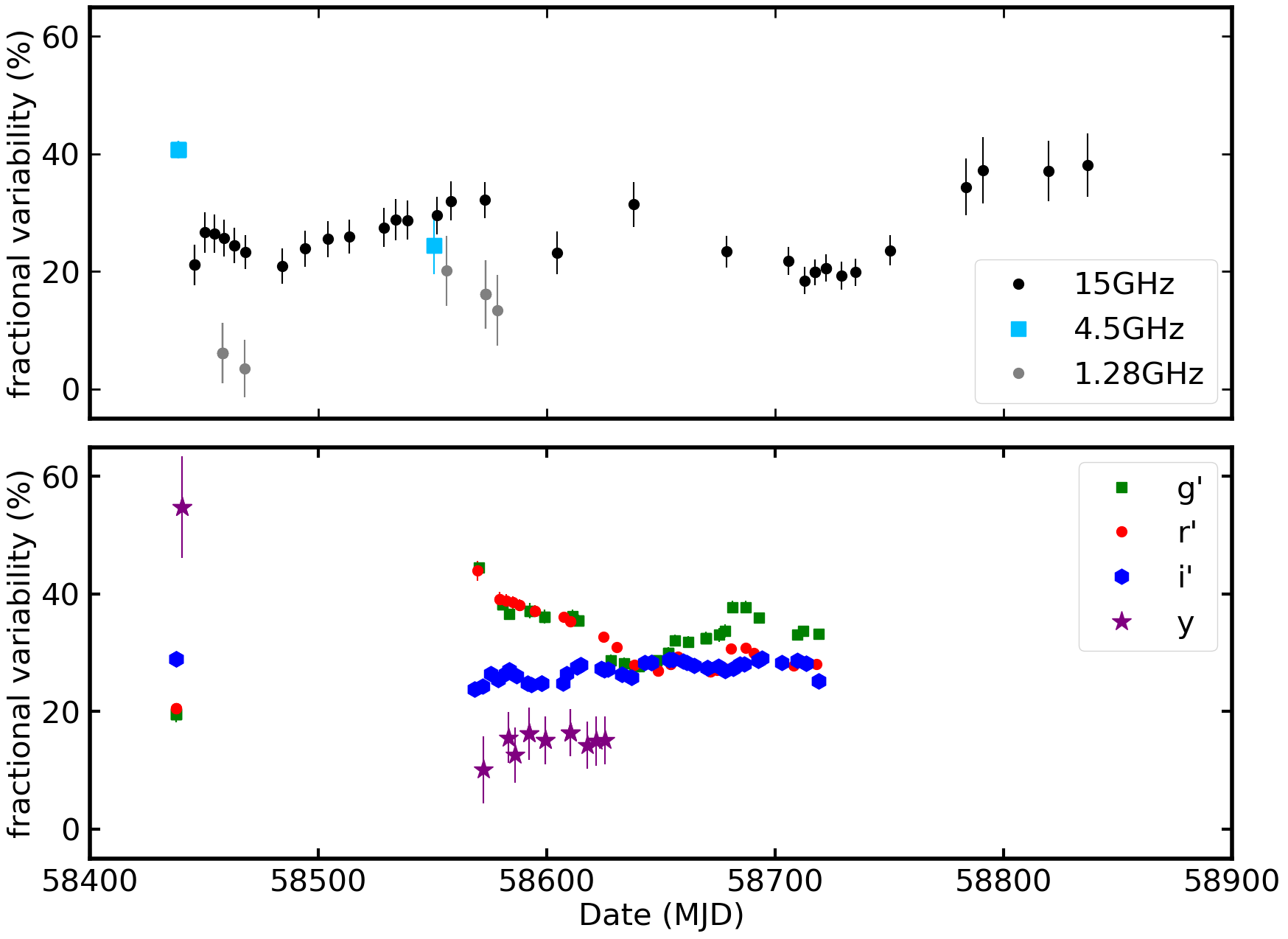}
    \caption{The radio (top panel) and optical (bottom panel) F\textsubscript{rms} variability as a function of time. We calculate a moving average rms variability with a bin size of 60 days. }
    \label{fig:rms_var_time}
\end{figure}

In addition to looking at the outburst-averaged variability properties, we also calculated a 60-day running average of the rms variability for each band. The results are shown in Figure \ref{fig:rms_var_time}. A 60-day bin size was used such that the only bins with fewer than 3 observations are those that were around periods of Sun-constraint. We found that reducing the bin size to 30 or 45 days amplified potential trends but also increased the uncertainties thus reducing the significance of any variations. Before J1858 was Sun-constrained, we find that the lower frequencies had larger fractional variability values ($y$ and $i^{\prime}$-bands). After Sun-constraint, this swaps with the highest frequency $g^{\prime}$ and $r^{\prime}$-bands demonstrating stronger variability. As the outburst progresses, the variability across all bands becomes very similar at around $30\%$. The 15\,GHz fractional variability smoothly varies between $\sim20$ and $\sim40\%$ through the outburst, the same range as measured in the $g^{\prime}$ and $r^{\prime}$ bands. The 1.3\,GHz values systematically sit lower, always below $20\%$ but with much larger uncertainties.


\begin{table*}
\caption{F\textsubscript{rms} variability values over the broadband spectrum of J1858, on various timescales, during the outburst. For the radio values we consider only the detections, and remove the milli-Jansky flare from the MeerKAT data set. For a sequence of observations, the frequency range probed (column 8) is the frequency corresponding to the total length of time of the sequence (minimum) and the time resolution (maximum).}            
\label{tab_rms}      
\centering                       
\begin{tabular}{l l l l l l l l l}     
\hline    
Regime & Waveband & Central $\nu$ (Hz)& Facility & MJD range & F\textsubscript{rms} (\%) & Cadence & Frequency range (Hz) & Reference \\
\hline
Radio &  L-band & $1.28 \times 10^{9}$ & MeerKAT & 58432--58700 & 20$\pm$4 & 7 d & $2.4 \times 10^{-8}$ -- $1.9 \times 10^{-6}$ & \cite{Rhodes2022} \\
Radio & C-band & $4.5 \times 10^{9}$ & VLA & 58431--58701 & 46$\pm$10 & 1 d &  $4.3 \times 10^{-8}$ -- $6.7 \times 10^{-5}$ & \cite{vandenEijnden2020} \\
Radio & Ku-band & $1.55 \times 10^{10}$ & AMI-LA & 58424--58886& 33$\pm$2 &  $\sim$7d & $2.5 \times 10^{-8}$ -- $1.1 \times 10^{-5}$  & \cite{Rhodes2022} \\
Optical & ${g}^{\prime }$-band & $6.29 \times 10^{14}$ & LCO & 58431--58791 & $35.5 \pm 0.6$ & $\sim$7d & $3.2 \times 10^{-8}$ -- $2.9 \times 10^{-4}$ & This work \\
Optical & ${r}^{\prime }$-band & $4.83 \times 10^{14}$ & LCO & 58431--58791 & $31.4 \pm 0.4$ & $\sim$11d & $2.7 \times 10^{-4}$ -- $2.1 \times 10^{-8}$ & This work \\
Optical & ${r}^{\prime }$-band & $4.83 \times 10^{14}$ & LCO & 58428 & $21.38 \pm 0.08$ & 46 s & $4.3 \times 10^{-4}$ -- $2.2 \times 10^{-2}$ & This work \\
Optical & ${i}^{\prime }$-band & $3.98 \times 10^{14}$ & LCO & 58431--58791 & $26.6 \pm 0.4$ & $\sim$5d & $4.7 \times 10^{-8}$ -- $2.4 \times 10^{-6}$ & This work \\
Optical & ${y}$-band & $2.99 \times 10^{14}$ & LCO & 58431--58791 & $19 \pm 3 $&  $\sim$8d & $4.5 \times 10^{-8}$ -- $1.0 \times 10^{-5}$ & This work \\

\hline                     
\end{tabular}
\end{table*}

\subsection{Spectral energy distributions}\label{Sec:SEDs}

\begin{figure}
    \centering
    \includegraphics[width = \columnwidth]{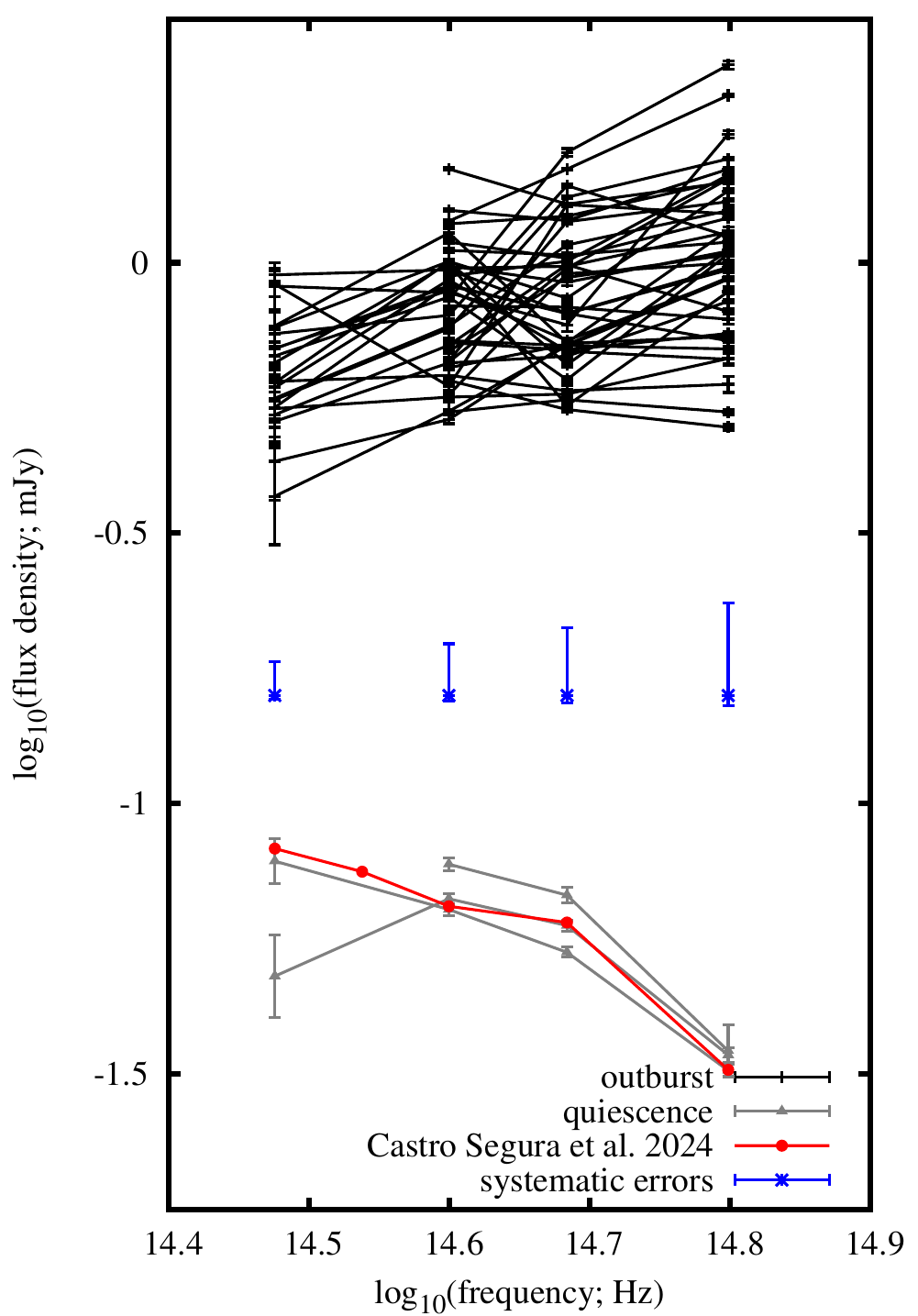}
    \caption{Optical SEDs of J1858 for all epochs where data were collected in at least three bands. The black, brighter data points correspond to when the source was in outburst. There are substantial changes in the spectral index during outburst. The fainter grey points are from observations taken during quiescence, and we add one SED from the quiescence period before the outburst, taken by Pan-STARRS \citep{CastroSegura2024}. The (blue) systematic errors represent how the SEDs would change with varying extinction. If $A_V = 1.0$ instead of the chosen $A_V = 0.64$ (see Section 3.2) the resulting shift in the data points across the four filters are denoted by the upper blue error bars.}
    \label{fig:opt_seds-all}
\end{figure}

For all epochs where observations were made in at least three bands on the same date, we construct optical SEDs, shown in Figure \ref{fig:opt_seds-all}. The optical magnitudes were de-reddened and converted to flux densities. Calculating the contribution of dust along the line of sight is complicated by strong, variable, intrinsic absorption that affects the soft X-rays, with measured neutral hydrogen columns of up to $N_H=$ (1.4 -- 4.2)$\times 10^{23}$\,cm\textsuperscript{-2} \citep{Hare2020}. However, during times when intrinsic absorption was not present, $N_H$ was found to be consistent with the Galactic value \citep[e.g.][]{Hare2020,CastroSegura2022,shabaz2023}. Using measurements from \citet{HI4PI2016}, we find that in a 5.5\,arcminute region about the source position the Galactic $N_H$ is $(1.84 \pm 0.03) \times 10^{21}$\,cm\textsuperscript{-2}. Absorption coefficients for J1858 were evaluated assuming this value and the \citet{Foight16} $N_{H}/A_{V}$ relation for our Galaxy, resulting in $A_{V}=0.64 \pm 0.04$. The optical data were de-reddened using the \citet{Cardelli1989} extinction law, which adopts extinction values (to be multiplied by $A_{V}$) of 1.194, 0.872, 0.666 and 0.438 in the $g'$, $r'$. $i'$ and $y$-bands, respectively. We note that \cite{CastroSegura2024} derived an extinction of $A_{V}=1.00 \pm 0.08$ by fitting the profile of the interstellar absorption feature near 2175\,\r{A} in the near-UV spectrum from HST. However, this would translate to \citep[using][]{Foight16} a neutral hydrogen column that is significantly higher than the Galactic value. It is possible that dust intrinsic to the system could contribute extra absorption at some orbital phases (see Section 4). For these two reasons, we adopt the Galactic absorption value in the below sections but note any changes that a higher absorption could make to the results. We also note that the much higher, intrinsic, variable neutral hydrogen column of $N_H > 1.4 \times 10^{23}$\,$cm^{-2}$ does not affect the optical emission. If the dust extinction were to correlate with this intrinsic $N_H$ according to the \cite{Foight16} law, the resulting in variable absorption would be $\Delta A_{V} > 48$ which is much higher than the observed amplitude. A similar result was found in the black hole XRB V404 Cyg, which also had strong intrinsic $N_H$ and no corresponding variable $A_{V}$ seen in optical and UV data \citep{Oates2019}.

For the data obtained during the outburst, we find that there is a large scatter in the spectral index measurements. Some of the individual SEDs show sharp peaks and troughs within the four bands, which is unphysical and is reflected in the large uncertainties on some of the spectral indices. This is a result of the large amplitude variability demonstrated in Figures  \ref{fig:g_vs_i} and \ref{fig:r-band-timing} where the source varies on timescales shorter than the time taken to switch between filters. The continuum emission cannot vary so abruptly within this small wavelength range. Prominent emission and absorption lines could, but they would have to exhibit very high amplitude variability and dominate over the continuum in all bands, which is not expected and such behaviour was not seen in optical spectra \citep{MunozDarias2020}. Combined with the lack of correlation between the magnitudes in different bands, it is not possible to measure long-term changes in the optical spectral index during the outburst. We can however estimate an average value for the optical spectral index, which follows $F_{\nu} \propto \nu^{0.67\pm0.09}$. This is a blue spectrum, which is fairly typical of low-mass X-ray binaries in outburst \citep[e.g.][]{hynes2005}, despite the atypical large amplitude short-term variability. The spectral index values range from $\alpha \sim 1.6$--1.8, which is near the Rayleigh-Jeans limit of a blackbody ($\alpha = 2.0$), to $\alpha \sim -0.6$-- -0.7, which is more consistent with optically thin synchrotron emission. We found that, with some scatter, some of the highest (most positive) spectral index values were when the $g^{\prime}$-band flux densities were brightest, and some of the lowest spectral index values were when the $g^{\prime}$-band flux densities were faintest as shown in Figure \ref{fig:cmdmodel}. 

We are also able to measure the spectral energy distribution during quiescence. Due to the contamination of the nearby interloper star 2.0\,arcseconds from J1858, we only include data on dates when solid detections were made (not forced photometry points) and when the seeing was $< 1.4$\,arcseconds. This ensured that the MAP photometry was able to exclude the flux from this nearby star from the aperture. 

During quiescence, we find that the spectral index is between $\alpha = -1$ and $-2$, which is consistent with the typical SED of a star or cold, quiescent disc. The SEDs are similar to one derived from Pan-STARRS magnitudes during quiescence before the outburst \citep{CastroSegura2024}, shown by red circles in Figure \ref{fig:opt_seds-all}.

\subsection{Colour magnitude diagram}\label{Sec:CMD}

\begin{figure}
    \centering
    \includegraphics[width = \columnwidth]{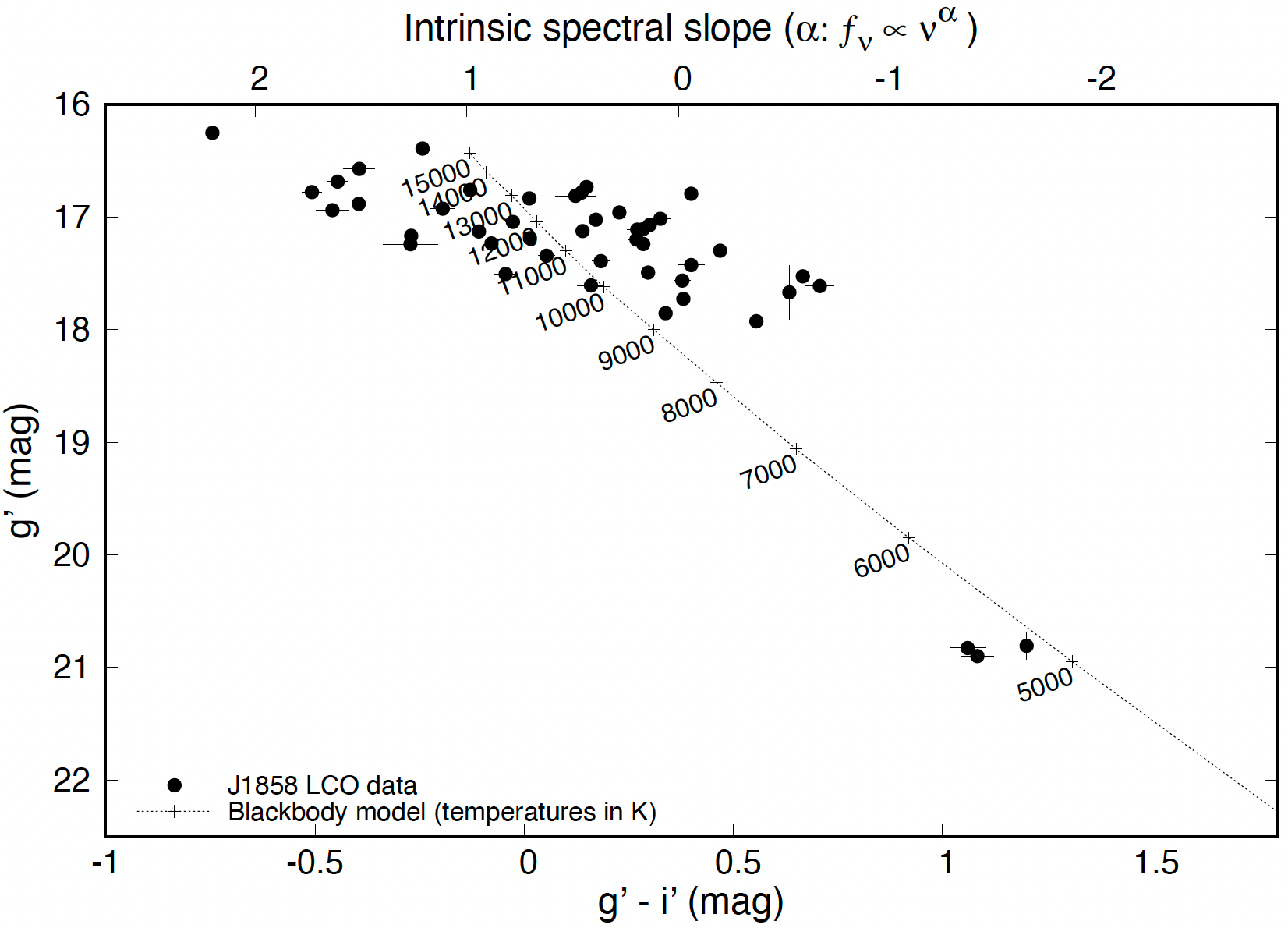}
    \caption{Colour magnitude diagram of J1858 showing optical brightness $g^{\prime}$ vs color $g^{\prime}$-$i^{\prime}$ where the bluer colours that correspond to higher spectral indices are shown to the left, and redder colors are shown to the right, overplotted with a simple model of a single temperature blackbody heating up and cooling denoted by the black dotted line labelled with temperature values.}
    \label{fig:cmdmodel}
\end{figure}

To analyse the colour evolution of J1858, we constructed a colour-magnitude diagram (CMD) using quasi-simultaneous (observations separated in time by less than $20$ minutes) $g^{\prime}$ and $i^{\prime}$-band LCO data (see Figure \ref{fig:cmdmodel}). We assume an optical extinction of $A_{V}=0.64$ (see Section \ref{Sec:SEDs}) to convert the $g^{\prime}$-$i^{\prime}$ color into an intrinsic spectral index. We find that when J1858 is fainter in quiescence, it populates the bottom-right corner of the CMD, when the emission could be dominated by the donor star. The source is much brighter and the colour bluer during the outburst, when the emission in LMXBs is expected to be dominated by the accretion disc. We superimpose a blackbody model \citep[][]{Maitra2008,Russell11} depicting the evolution of a single-temperature, constant-area blackbody that has a varying temperature in Figure \ref{fig:cmdmodel} (the grey line labelled with temperatures in Kelvin). The normalisation of the blackbody model is dependent on the projected surface area of the disc and its luminosity whose parameters are then dependent on several parameters including the accretion disc radius, disc filling factor, distance and orbital period. The latter two, among others, are well constrained \citep{2020MNRAS.499..793B, 2021MNRAS.503.5600B} whereas others are not and are estimated by marginalizing over when comparing the blackbody model to the data \citep[][]{Zhang2019,baglio2020,2022ApJ...930...20B,saikia1716,saikia2023}. The temperature corresponding to the optical colours is $\sim$5,000 K during quiescence \citep[which is close to that of the donor star, derived in][]{CastroSegura2024} and $\sim$8,000--15,000 K during outburst, consistent with an ionized disc \citep[hydrogen is completely ionized above 10,000 K; e.g.][]{Lasota2001} or the dayside of an irradiated companion star \citep{2009MNRAS.399.2055B}.

We find that the slope of the model does not approximate the slope in the trend of the data in outburst well. This cannot be due to uncertainty in our optical extinction value (changing $A_V = 0.64$ to $A_V=1.0$ would make the SEDs slightly bluer but would not be able to account for the shallow slope in the data compared to the model). The shallow slope could imply that a standard thermal (irradiated) disc is not producing all of the optical emission during the outburst, or that the fast variability could be responsible. Since the $g'$ and $i'$-bands are only mildly correlated (Figure 2 ), the scatter in the $g'-i'$ colour ($\sim1.5$\,mag) will be $\sim \sqrt2$ times the scatter in each band, spreading the data out more than the blackbody model predicts in the x-axis of Figure \ref{fig:cmdmodel}. 
One possible explanation is that an irradiated disc produces the optical emission, but varies on short timescales due to a highly variable irradiating X-ray flux \citep[as discussed in Sections \ref{subsec:long_term_optical} and \ref{sec:discussion} and seen by][]{shabaz2023}, leading to scatter in the colour around the mean disc value. This results in the model being a poor approximation of the slope of the trend in the data, although, the model (normalized to the centre of the scattered points) does pass through the quiescent data points which is either a coincidence or lends itself to the possibility of a the disc still contributing in quiescence.

\subsection{Periodicity analysis}

\begin{figure}
    \centering
    \includegraphics[width = \columnwidth]{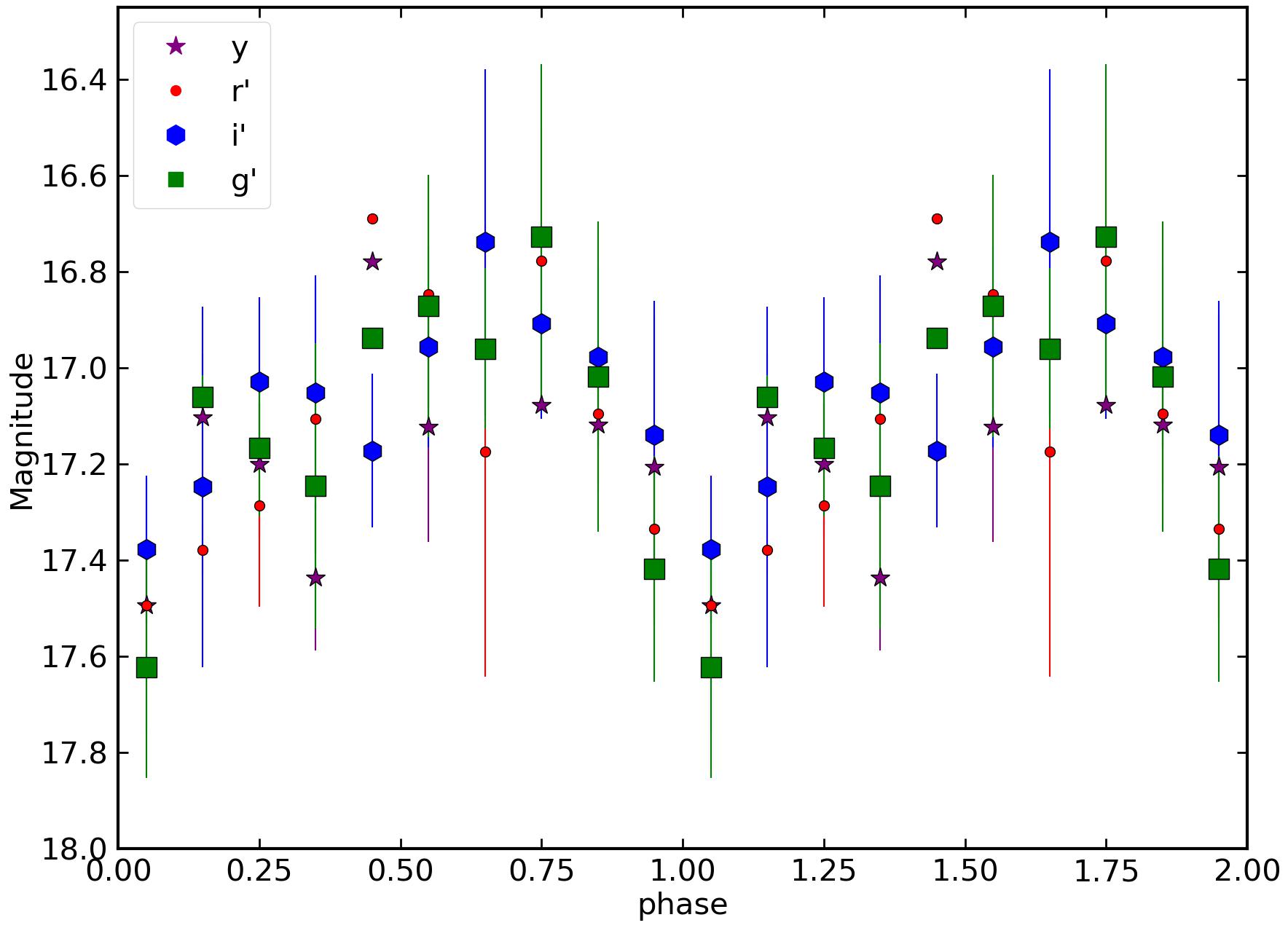}
    \caption{Phase folded light curve for our optical data during the outburst. Each phase bin is 10\% of the phase. The uncertainties on each data point reflect the scatter within that phase bin. Points with no error bars only have one observation within that phase bin.}
    \label{fig:opt_phase}
\end{figure}

Given the temporal extent of our monitoring campaign, we folded the optical light curve on the orbital period of the binary: (21.3448$\pm$0.0004)\,hours to search for any signs of periodicity \citep{2021MNRAS.503.5600B}. We denote phase ($\phi$) of 0.0 as the point at superior conjunction of the NS to be consistent with other publications \citep[e.g.][superior conjunction of the NS refers to when the companion star is directly between the observer and the NS]{2022MNRAS.514.1908K}. We plot the phase curve for all optical bands (with no correction for extinction) in Figure \ref{fig:opt_phase} and find evidence for phase dependence in all bands. The optical counterpart is brightest between phases of 0.6 and 0.8 and faintest at 0.0 when the NS is being eclipsed. We find that the amplitude modulation is the same ($\sim$0.7\,magnitudes) across all bands indicating that a geometry change in the system is causing the phase-dependent changes rather than extinction. This is different to the rms variability measurements which find that the shortest wavelength bands ($g^{\prime}$ and $r^{\prime}$) have the largest rms variability amplitude which is seen in the large uncertainties on each phase bin.

\begin{figure}
    \centering
    \includegraphics[width=\columnwidth]{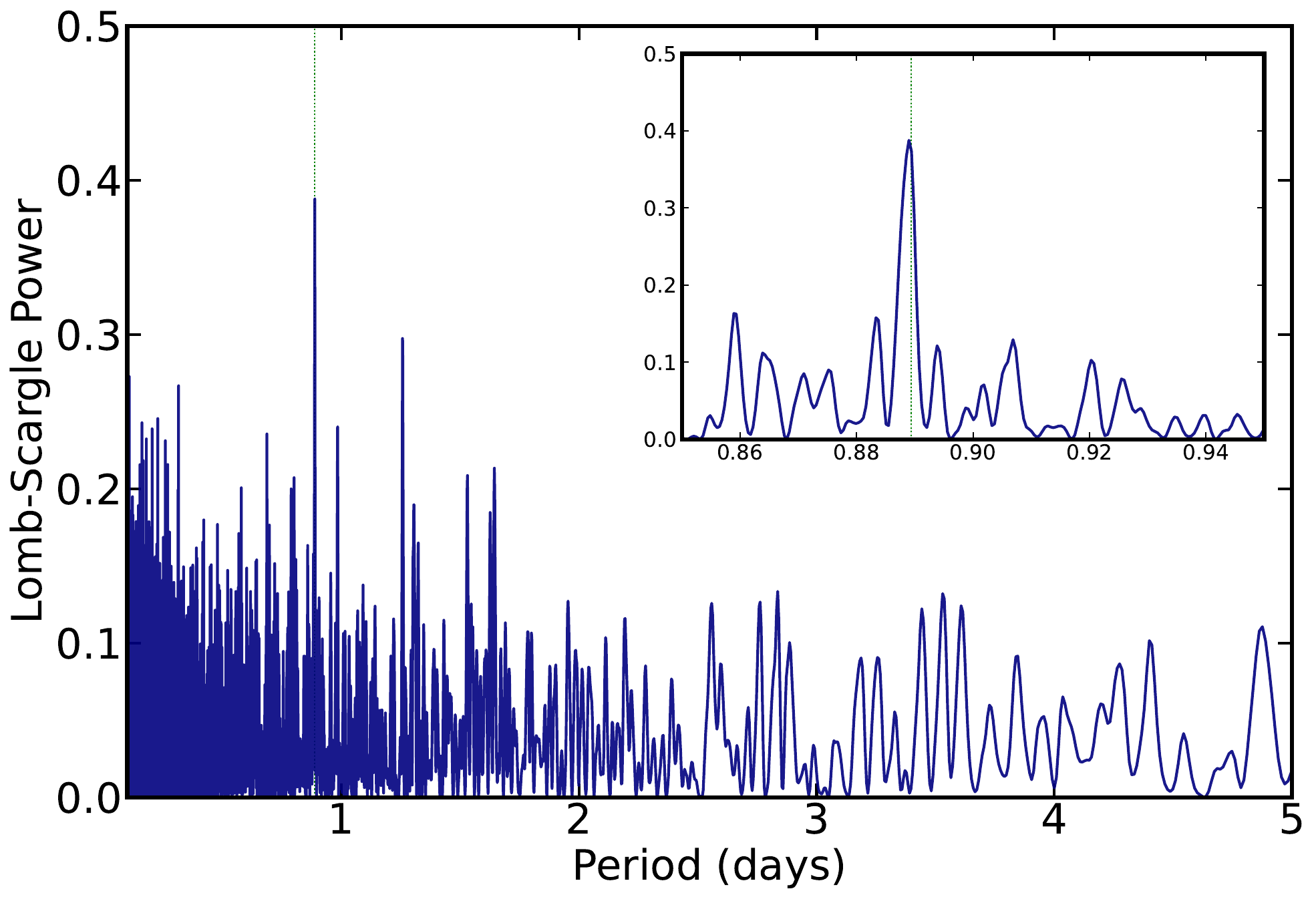}
    \caption{A Lomb-Scargle Periodogram for the optical light curve data for J1858. Between periods of 0.1 and 5.0\,days, we find the strongest peak at 0.889\,days with an estimated peak width of 0.001\,days \citep{2015ApJ...812...18V, jake_vanderplas_2015_14833}. This result is consistent with the period derived from eclipse timings in \citet{2021MNRAS.503.5600B} denoted with a vertical blue line at 0.88937\,days.}
    \label{fig:lombscargle}
\end{figure}

To ensure that the behaviour we see in Figure \ref{fig:opt_phase} is real, we performed an F-test comparing a flat phase curve and a skewed sinusoid \citep{1779952}. We find that the skewed sinusoid model is preferred at a significance level of $99.5\%$ level. We also constructed a Lomb-Scargle Periodogram to search for periodic behaviour in the optical light curve with periods between 0.1 and 5\,days \citep{2015ApJ...812...18V, jake_vanderplas_2015_14833}. The Lomb-Scargle Periodogram is shown in Figure \ref{fig:lombscargle} with a vertical line denoting the orbital period as measured from the eclipses in \citet{2021MNRAS.503.5600B}. We find a peak at 0.889$\pm$0.001\,days which is consistent with the measured orbital period of J1858 which is 0.88937$\pm$0.00002\,days. We note that the orbital period was not given as a prior in the Lomb-Scargle Periodogram, which, when combined with the F-test results leads us to conclude that the phase dependence is real.

\section{Discussion}
\label{sec:discussion}

\begin{figure*}
    \centering
    \includegraphics[width=0.8\textwidth]{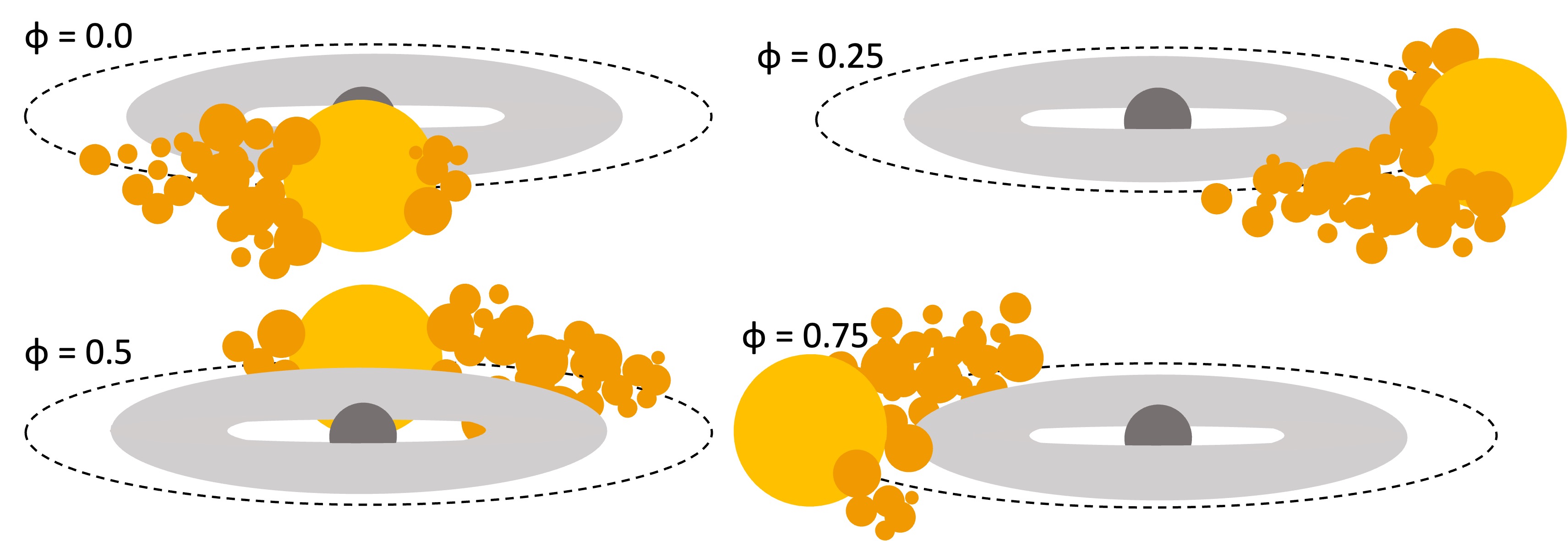}
    \caption{Schematic of how the phase curve (Figure \ref{fig:opt_phase}) is produced. At phase 0.7, the projected surface area of the star and ablated material is larger and therefore the reprocessed radiation along the observer's line of sight is larger resulting in a higher flux density. Whereas at phase 0.0 the companion star completely eclipses the NS and so all reprocessed radiation from the star and ablated material travels away from the observer.}
    \label{fig:model}
\end{figure*}

Our long-term optical monitoring campaign of J1858 has shown significant variability through the outburst (see Figures \ref{fig:optical_radio_lightcurve}, \ref{fig:r-band-timing} and \ref{fig:rms_var_time}). The simplest explanation for the optical behaviour that we observe is that we are randomly sampling the short-timescale rapid flaring similar to that in Figure \ref{fig:r-band-timing} and has been reported elsewhere \citep[][]{Vincentelli2023, shabaz2023}. We compare our long-term data set with the short-term study from \citet{shabaz2023} to test this hypothesis. We find that the average magnitude measurements in the $g^{\prime}$, $r^{\prime}$ and $i^{\prime}$-bands are fully consistent however comparing the levels of variability between the short (minute) timescale studies from \citet{shabaz2023} to our longer (weeks-months) values, we measure slightly lower variability values. In $g^{\prime}$, $r^{\prime}$ and $i^{\prime}$-bands, \citet{shabaz2023} measure F\textsubscript{rms} values of 27, 22 and 23\% compared to our 22, 20 and 17\%.

\citet{Vincentelli2023} suggests that the variability is a result of changes in mass accretion rate i.e. the system is constantly going through a pattern of ejecting and refilling the inner accretion disc \citep{1997ApJ...479L.145B}. This scenario is expected to result in the repeated ejection of plasmoids each producing radio emission that transitions from optically thick to thin as they expand. High-time-resolution radio observations are needed to resolve individual ejecta which evolve on timescales of minutes \citep{vandenEijnden2020, Vincentelli2023, Rhodes2022}. In \citet{Rhodes2022} and \citet{Vincentelli2023}, high-time-resolution radio observations showed a highly variable radio source that transitioned from optically thick to thin over timescales of minutes as would be expected for repeated rapid ejections. Observations that lasted longer than 15\,minutes were averaging over multiple evolving plasmoids producing a flat/optically thick variable radio source, as is shown in the lower panel of Figure \ref{fig:optical_radio_lightcurve}. Therefore, although we cannot directly correlate the behaviour between optical and radio bands, we can connect the variability observed at optical wavelengths due to a varying accretion rate to that at radio frequencies due to changing jet power that continues throughout the outburst.

The above statements assume that the jet dominates emission only at radio frequencies, however depending on the relative strength of different ejecta, the contribution at higher frequencies may begin to dominate over the emission from the disc or irradiated stellar companion. The optical spectral index, if jet/ejecta dominated, would follow $F_{\nu} \propto \nu^{<0}$. We measure negative spectral indices in eight observations during the outburst. The flux densities of the optical emission are high enough ($\sim$1\,mJy) that it would be possible that the optical and radio emission (100s $\mu$Jy with an optically thick spectral index) could originate from the same process i.e. synchrotron emission from the jet. Therefore, we find it likely that there is also a jet contribution to the optical variability.

One of the most interesting discoveries from this observing campaign is the apparent phase dependence in the optical light curve data (Figure \ref{fig:opt_phase}). A possible interpretation of the phase-dependent behaviour is a superhump, which is thought to occur due to precession in the disc \citep{1991MNRAS.249...25W}. The period of the superhump is expected to be close to or a few per cent different to the orbital period and so could be visible in the orbital phase curve. We rule out the possibility of a superhump based on the maximum difference between the brightest and faintest points in the phase curves shown in Figure \ref{fig:opt_phase} being about 0.7\,magnitudes. This is larger than the largest modulation amplitude (0.5\,magnitudes) found to date caused by a superhump which was found in MAXI J1820-070, a BH LMXB \citep{2022MNRAS.513L..35T}. Furthermore, the period of a superhump is expected to evolve as the outburst progresses which we find no evidence of in J1858, thus making a superhump even less likely as the origin of the phase variability.

Instead, we interpret the orbital modulation as a result of irradiation within the system. From \citet{2021MNRAS.503.5600B, 2022MNRAS.514.1908K}, we know that the system has a high inclination (i.e. is almost edge-on) and given the results from \citet{Vincentelli2023} and \citet{2023MNRAS.520.3416K}, where strong irradiation is inferred, one would expect the optical phase curve to be sinusoidal. However, the observed phase curve deviates from a sinusoid with the peak occurring at $\phi \approx 0.7$ rather than 0.5 and the minimum occurring at $\phi \approx 0.0$. We interpret the asymmetry as the result of additional material gravitationally bound to the system. The material is thought to be removed from the stellar companion as a result of high energy radiation from the inner accretion flow irradiating the companion. Extended material has been invoked in radio and X-ray observations from spider pulsars and other X-ray binaries, respectively \citep{1988Natur.333..237F, 2023MNRAS.520.3416K}.

To determine how the ablated material is changing the phase curve, we compare the amplitude of the modulation in Figure \ref{fig:opt_phase} across all four wavebands. We find no colour dependence in the phase curve shape: the difference between the peak and trough across all bands is about 0.7\,magnitudes. If the shape was due to dust absorption of light by ablated material, we would expect the amplitude of the phase curve to be 2.7 times greater at g' than y band. No such dependence is observed. A flat frequency dependence could be a result of complete obscuration however the magnitude decrease around phase 0.0 is not sufficient for complete obscuration. 

Instead, we hypothesise that the peak in the phase curve is caused by increased projected surface area for reprocessing of the X-ray emission from the inner accretion disc into our line of sight. A schematic of this scenario is shown in Figure \ref{fig:model}. At $\phi = 0.0$, the NS is being eclipsed by the companion star, the disc is not completely obscured by the companion/ablated material which is why the flux level at phase 0.0 does not drop as low as in quiescence. Because the compassion star is between us and the neutron star, we do not observe any reprocessed emission. Between phase 0.0 to $\sim$0.6, the sky-projected area available for reprocessing hard X-ray emission into our line of sight slowly increases to a maximum around 0.7 where the ablated material nor the star is blocked by the disc. The system appears brightest when the surface area is largest and so there is more X-ray radiation being reprocessed into our line of sight (see $\phi = 0.75$ in Figure \ref{fig:model}). Between phases 0.7 and 1.0, we observe less reprocessed material because the apparent size of the star and ablated material is smaller as the material sits behind the star according to our viewing angle and so the system appears fainter.

Phase folding of X-ray binary optical light curves is usually performed when the source is in quiescence in order to study the ellipsoidal modulation of the companion \citep{1975ApJ...197..675A}. The behaviour we observed cannot be explained by ellipsoidal modulation. Instead, the behaviour we observe here is very similar to that observed in transitional millisecond pulsars \citep[e.g.][]{2021MNRAS.507.2174S, 2024A&A...691A..36D}. The phase dependence is often concluded to be due to irradiation of the companion with no need for ablation to explain the data. Evidence for ablation in neutron star systems was first found in eclipsing millisecond pulsars called spider pulsars where there is an increased density of material around the ingress and egress of the pulsar eclipse \citep{1988Natur.333..237F, 2019MNRAS.490..889P}, and more recently in neutron star X-ray binaries \citep{2023MNRAS.520.3416K}. In the case of spider pulsars, the mass loss rates inferred from the ablation will only just, or is insufficient, to fully evaporate the stellar companion within a Hubble time which creates problems if spider pulsars are the progenitors of isolated millisecond pulsars \cite{2020MNRAS.494.2948P, 2020MNRAS.495.3656G}. Our observations, which we interpret as ablated material extended around the binary orbit throughout the outburst, provide a possible path to increase the total time over which evaporation of the companion can occur and therefore make isolated millisecond pulsars easier to produce.

\section{Summary and conclusions} \label{sec:conclusions}

In this paper, we have presented a long-term optical monitoring campaign for the NS LMXB \textit{Swift} J1858.6--0814 both in outburst and after in quiescence. Our observations show significant frequency-dependent variability throughout the outburst. We compare variability observed in the weekly optical and radio observations \citep[the latter from][]{vandenEijnden2020, Rhodes2022} and find that the variability amplitude is similar indicating that the cause of the variability is likely to be a common process in the two bands: a result of changes in mass accretion rate. We have also shown that there are processes ongoing within the J1858 system that affect the optical emission on other timescales: the jet flares that last less than 15\,minutes \citep{Rhodes2022, Vincentelli2023} and phase-dependent behaviour on timescales of hours and so the lack of a good fit for a single black body emitting region as shown in Figure \ref{fig:cmdmodel} is expected.

By folding the optical data about the orbital period, we find significant orbital modulation which disappears in quiescence. Such behaviour is reminiscent of spider pulsars where the pulsar wind ablates material from the surface of the stellar companion and reduces its mass. However, unlike in spider pulsars, our observations find evidence for ablation while the NS system is actively accreting as an LMXB thus providing a longer timescale over which a stellar companion could be destroyed and leave behind an isolated MSP.

\section*{Acknowledgements}

We thank the anonymous referee for their help in improving the manuscript. We thank Dan Bramich for his contributions to the development of the XB-NEWS pipeline, and for insightful discussions.
AHK acknowledges support from the Science and Technology Facilities Council (STFC) as part of the consolidated grant award ST/X001075/1. JvdE acknowledges a Warwick Astrophysics prize post-doctoral fellowship made possible thanks to a generous philanthropic donation.
This work uses data from the Faulkes Telescope Project, which is an education partner of Las Cumbres Observatory (LCO). The Faulkes Telescopes are maintained and operated by LCO. 
This material is based upon work supported by Tamkeen under the NYU Abu Dhabi Research Institute grant CASS. MCB acknowledges support from the INAF-Astrofit fellowship.

\section*{Data Availability}

The optical data from Faulkes/LCO underlying this article will be shared on reasonable request to the corresponding author.



\bibliographystyle{mnras}
\bibliography{bib} 



\appendix


\bsp	
\label{lastpage}
\end{document}